\documentclass[aps,nofootinbib,prd,twocolumn,preprintnumbers]{revtex4-1}
\pdfoutput=1 % if your are submitting a pdflatex (i.e. if you have
\usepackage[T1]{fontenc} % if needed
\usepackage{graphicx}
\usepackage{epsfig}
\usepackage{rotating}
\usepackage{color}
\usepackage[normalem]{ulem}
\usepackage{multirow}
\usepackage{tabularx}
\usepackage{graphicx}
\usepackage{dcolumn}
\usepackage{hyperref}
\usepackage{float}
\usepackage{amsmath}
\usepackage{amssymb}
\usepackage{float}

\usepackage{comment}
\def\vev#1{\left\langle #1\right\rangle}

\excludecomment{ignore}

\usepackage{ulem}
\usepackage{color}

\newcommand{\AddrAHEP}{
  {\it AHEP Group, Instituto de F\'{\i}sica Corpuscular --
    C.S.I.C./Universitat de Val{\`e}ncia \\
    Edificio de Institutos de Paterna,
 C/Catedratico Jos\'e Beltr\'an, 2 E-46980 Paterna (Val\`{e}ncia) - Spain}}

\begin{document}

%\title{\red{Signatures of light sterile neutrinos  at reactor \\ and Spallation
%  Neutron Source neutrino experiments} }
\title{Probing light sterile neutrino signatures at reactor and Spallation Neutron Source neutrino experiments}
\author{T.S. Kosmas~$^1$}\email{hkosmas@uoi.gr}
\author{D.K. Papoulias~$^{1}$}\email{dimpap@cc.uoi.gr}
\author{M. T\'ortola~$^2$}\email{mariam@ific.uv.es}
\author{J.W.F. Valle~$^2$}\email{valle@ific.uv.es, http://astroparticles.es/} 

\affiliation{$^1$~Theoretical Physics Section, University of Ioannina,
  GR-45110 Ioannina, Greece} \affiliation{$^2$~\AddrAHEP}

\begin{abstract} 

  We investigate the impact of a fourth sterile neutrino at reactor
  and Spallation Neutron Source neutrino detectors. Specifically, we
  explore the discovery potential of the TEXONO and COHERENT
  experiments to subleading sterile neutrino effects through the
  measurement of the coherent elastic neutrino-nucleus scattering
  event rate. Our dedicated $\chi^2$-sensitivity analysis employs
  realistic nuclear structure calculations adequate for high purity
  sub-keV threshold Germanium detectors.

\end{abstract}
%\pacs{13.15.+g,14.60.St,12.60.-i,13.40.Em} 

\maketitle
%\flushbottom

% \linenumbers

\section{Introduction}
\label{sec:intro}

Recently, several neutrino experiments have been designed to operate
with exceptional high sensitivities in order to detect neutral-current
coherent elastic neutrino-nucleus scattering (CE$\nu$NS)
events~\cite{Scholberg:2005qs,Wong:2010zzc}
for the first time~\cite{Brice:2013fwa,Collar:2014lya}.
Potential deviations from the standard model (SM) expectations would
provide a glimpse on new
physics~\cite{Papoulias:2013gha,Papoulias:2015iga,Kosmas:2015sqa}.
Indeed, the existence of a fourth sterile neutrino could be probed in
ultralow threshold neutrino-nucleus coherent scattering, since it
would generate tiny modifications in the final neutrino
spectrum~\cite{Anderson:2012pn,Dutta:2015nlo}. The purely neutral
character of CE$\nu$NS provides an important advantage~\cite{Freedman:1973yd,Tubbs:1975jx,Drukier:1983gj}, compared to
neutrino-electron scattering since there is no need for disentangling
the sterile neutrino mixing from that of the active
neutrinos~\cite{Formaggio:2011jt}.

On the other hand the solid evidence for neutrino oscillations implied by current solar
and atmospheric data, and confirmed by reactor and accelerator neutrino
experiments~\cite{Kajita:2016cak,McDonald:2016ixn,Forero:2014bxa,Valle:2015pba}
still leaves some loopholes.
These come in the form of controversial anomalies which do not fit in
the three-neutrino oscillation paradigm.
The Gallium~\cite{Hampel:1997fc,Abdurashitov:2009tn},
LSND~\cite{Athanassopoulos:1995iw,Aguilar:2001ty}, and
MiniBooNE~\cite{AguilarArevalo:2007it,Aguilar-Arevalo:2012fmn,Aguilar-Arevalo:2013pmq} anomalies, as well as the new
predictions for reactor neutrino
fluxes~\cite{Mention:2011rk,Mueller:2011nm,Huber:2011wv} have raised
speculations on whether the actual number of neutrinos could exceed
three.
Taken at face value, these have suggested the possible existence of at
least one sterile neutrino with new mixings to the three active
neutrinos.
The indicated squared mass splittings are of the order of
 $1 \, $eV$^2$~\cite{GonzalezGarcia:2007ib,Gariazzo:2015rra}.
Following earlier
theoretical~\cite{peltoniemi:1993ec,peltoniemi:1993ss} and
phenomenological considerations~\cite{Maltoni:2004ei}, the possible
existence of a fourth neutrino has drawn a lot of attention and many
recent studies have been carried
out~\cite{deGouvea:2008qk,Giunti:2011gz,Kopp:2013vaa,An:2014bik,Giunti:2015mwa}. In
fact, an arbitrary number of {$\mathrm{SU(2)_L}$ singlet fermions are
  present in the generalized type I seesaw
  mechanism~\cite{Schechter:1980gr} such as realized in low-scale
  seesaw schemes
  ~\cite{Mohapatra:1986bd,Akhmedov:1995ip,Akhmedov:1995vm,Malinsky:2005bi}.
  If it exists, the sterile neutrino is expected to take part in
  neutrino oscillations. Notice however that, despite the limits on
  the number of sterile neutrino states coming from
  cosmology~\cite{pastor-book}, depending on the active-sterile mixing
  strength and their corresponding mass scale, such cosmological
  constraints may be adequately fulfilled~\cite{Giunti:2015wnd}.
  Furthermore, sterile neutrino states may induce a number of
  processes with important phenomenological consequences to
  solar~\cite{deHolanda:2010am}, reactor~\cite{An:2013zwz,Ahn:2012nd,Abe:2011fz} and
  accelerator~\cite{Bandyopadhyay:2007kx} neutrino oscillations at the
  sub-eV scale, possible neutrino electromagnetic interactions at the
  eV scale~\cite{Balantekin:2013sda}, dark matter at the keV
  scale~\cite{Ando:2010ye,Liao:2013jwa}, etc. Moreover, the impact of
  a light sterile neutrino on the neutrinoless double beta-decay and
  single beta-decay processes has also received some
  attention~\cite{Bilenky:2002aw,Vergados:2012xy,Giunti:2010wz}. 

Here we examine the possibility of probing light sterile neutrinos at
short-baseline CE$\nu$NS experiments operating with nuclear detectors of
low-threshold
capabilities~\cite{Soma:2014zgm,Dutta:2015vwa,Chen:2016lyr}.
A number of experiments are now planned in order to probe possible
oscillation features due to the presence of sterile neutrinos.
Specifically we examine the observation potential of the COHERENT
experiment at Oak Ridge~\cite{Akimov:2015nza} and the TEXONO
experiment in Taiwan~\cite{Kerman:2016jqp,Sevda:2016otj}.  Other
relevant projects looking for this signature are the
$\nu$GeN~\cite{Belov:2015ufh} and the GEMMA~\cite{Beda:2013mta}
experiments in Russia, as well as the CONNIE project in
Brazil~\cite{Moroni:2014wia,Aguilar-Arevalo:2016qen} and the MINER
experiment at Texas A\&M University~\cite{Agnolet:2016zir}. Notable
efforts aiming at observing CE$\nu$NS by using cryogenic detector
techniques include the Ricochet~\cite{Billard:2016giu} and the
$\nu$-cleus~\cite{Strauss:2017cuu} experiments.
Our calculations are performed using advanced nuclear physics
techniques, such as the quasiparticle random phase approximation
(QRPA), in which the required nuclear form factors are obtained with
high accuracy~\cite{Papoulias:2015vxa}. We also address the quenching
effects which are crucial in order to provide realistic
results~\cite{Kosmas:2015vsa}.
For the specific case of the aforementioned reactor and spallation
neutron source (SNS) experiments, we perform a $\chi^2$ sensitivity
analysis to explore the possibility that the detection of
CE$\nu$NS~\cite{Lindner:2016wff,Dent:2016wcr,Coloma:2017egw} constitutes an efficient probe for sterile neutrino searches at low energies. 

The paper has been organized as follows. We first go through a brief
description of the relevant formalism of CE$\nu$NS including sterile
neutrinos in Sec.~\ref{sect:cenns}. In Sec.~\ref{sect:experiments} we
summarise the main features of the relevant experiments, such as TEXONO
and COHERENT, necessary for our work. In
Sec.~\ref{sec:impact-light-sterile} we discuss the impact of a light
sterile neutrino in neutrino-nucleus scattering.  The results of our
calculations are discussed in Sec.~\ref{sect:numer_results}, where we
extract the expected sensitivities on the model parameters. Finally,
in Sec.~\ref{sect:conclusions} we close with a summary of our main
conclusions.

\section{Coherent elastic neutrino-nucleus scattering}
\label{sect:cenns}

At low and intermediate energies, considered in the present study, the
neutral-current neutrino-nucleus processes are described by the matrix
elements of an effective interaction Hamiltonian, written in terms of
the leptonic $\hat{j}_{\mu}^{lept}$ and hadronic (nuclear)
$\hat{\mathcal{J}}^{\mu}$ currents as
\begin{equation}
\label{nucl-tran-ME}
\langle f|\hat{H}_{eff}|i\rangle = \frac{G_F}{\sqrt{2}} \int d^{3}\mathbf{x} \, \langle \ell_f \vert \hat{j}_{\mu}^{lept} \vert \ell_i\rangle \langle J_f \vert \hat{\mathcal{J}}^{\mu}(\mathbf{x}) \vert J_i\rangle \, ,
\end{equation}
where $G_F$ is the Fermi constant.  The matrix element of the leptonic
current, between an initial $\vert \ell_i\rangle$ and a final lepton
state $\vert \ell_f\rangle$ takes the usual V-A form
\begin{equation}
\langle \ell_f \vert \hat{j}_{\mu}^{lept} \vert \ell_i\rangle = \bar{\nu}_{\alpha} \gamma_{\mu}(1-\gamma_5) \nu_{\alpha}\,e^{-i{\mathbf{q} \cdot \mathbf{x}}}\, ,
\end{equation}
with $\alpha=\{e,\mu,\tau \}$ being the neutrino flavor and
$\mathbf{q}$ denoting the three momentum transfer. The hadronic matrix
element is obtained through a multipole decomposition as described in
Refs.~\cite{Giannaka:2015sta,Chasioti2009234}.  Then, the differential
cross section with respect to the scattering angle $\theta$, for the
CE$\nu$NS ($gs\rightarrow gs$ transitions) off a spherical spin-zero
nucleus, reads~\cite{Papoulias:2013gha,Papoulias:2015vxa}
\begin{equation}
\left( \frac{d\sigma}{d\cos \theta} \right)_{\mathrm{SM}} = \frac{G_F^2}{2 \pi} E_{\nu}^2 \left(1 + \cos \theta 
\right) \left\vert\langle gs \vert\vert \hat{\mathcal{M}}_{00}(Q) \vert\vert gs \rangle\right \vert^2\, .
\label{SM_dT}
\end{equation}
The coherent nuclear matrix element is written in terms of the left- and right-handed couplings of the $u$- and $d$-quarks to the $Z$-boson as~\cite{Papoulias:2013gha}
\begin{equation}
\begin{aligned}
& \left\vert\langle gs \vert\vert \hat{\mathcal{M}}_{00}(Q) \vert\vert gs \rangle\right \vert = \int d^3r \, j_0(\vert \mathbf{q}\vert r) \\
  \times & \Big\{  \left[ 2(g_{\alpha \alpha}^{u,L} + g_{\alpha \alpha}^{u,R}) + (g_{\alpha \alpha}^{d,L} + g_{\alpha \alpha}^{d,R}) \right]   \rho_{p}(r) \\
& +  \left[ (g_{\alpha \alpha}^{u,L} + g_{\alpha \alpha}^{u,R}) +2(g_{\alpha \alpha}^{d,L} + g_{\alpha
  \alpha}^{d,R})\right]   \rho_{n}(r)  \Big \} \, ,
  \end{aligned}
\label{SM-ME}
\end{equation} 
where the notation $r = \vert \mathbf{x}\vert$ has been
introduced. In the latter expression, $ \rho_{p}(r)$ and
$ \rho_{n}(r)$ are the corresponding proton and neutron charge density
distributions computed through realistic nuclear structure
calculations in the context of the QRPA method. In such calculations,
the finite nucleon and nuclear size are taken into consideration by
weighting the differential cross section with corrections provided by
the associated proton (neutron) nuclear form factors $F_{Z(N)}(Q^2)$
that depend on the square of the four momentum transfer
\begin{equation} 
-q_\mu q^\mu=Q^2 = 2 E_\nu^2 (1-\cos \theta)\,,
\label{eq:mom_transf}
\end{equation}
or $Q = 2 E_\nu \sin (\theta/2)$.     
In Eq.(\ref{SM-ME}), the $u$- and $d$-quark couplings to the $Z$-boson include the relevant radiative corrections, through the expressions
\begin{equation}
\begin{aligned}
g_{\alpha \alpha}^{u,L} =& \rho_{\nu N}^{NC} \left( \frac{1}{2}-\frac{2}{3} \hat{\kappa}_{\nu N} \hat{s}^2_Z \right) + \lambda^{u,L} \, ,\\
g_{\alpha \alpha}^{d,L} =& \rho_{\nu N}^{NC} \left( -\frac{1}{2}+\frac{1}{3} \hat{\kappa}_{\nu N} \hat{s}^2_Z \right) + \lambda^{d,L} \, ,\\
g_{\alpha \alpha}^{u,R} =& \rho_{\nu N}^{NC} \left(-\frac{2}{3} \hat{\kappa}_{\nu N} \hat{s}^2_Z \right) + \lambda^{u,R} \, ,\\
g_{\alpha \alpha}^{d,R} =& \rho_{\nu N}^{NC} \left(\frac{1}{3} \hat{\kappa}_{\nu N} \hat{s}^2_Z \right) + \lambda^{d,R} \, ,
\end{aligned}
\end{equation}
with $\hat{s}^2_Z = \sin^2 \theta_W= 0.23120$, $\rho_{\nu N}^{NC} = 1.0086$, $\hat{\kappa}_{\nu N} = 0.9978$, $\lambda^{u,L} = -0.0031$, $\lambda^{d,L} = -0.0025$ and $\lambda^{d,R} =2\lambda^{u,R} = 7.5 \times 10^{-5}$~\cite{Beringer:1900zz}.

\subsection{Nuclear physics calculations}
\label{sec:nucl-phys-calc}

It can be noticed that the CE$\nu$NS cross section is rather sensitive to
the neutron form factor, calculable in the context of a nuclear
structure model. In this work, the reliability of the evaluated cross
sections is maximized by performing QRPA calculations, incorporating
realistic strong nuclear forces within the framework of a
comprehensive phenomenological meson-exchange theory for the reliable
description of the nucleon-nucleon interaction. Our QRPA code, for the two-nucleon 
residual interaction utilizes the C-D version of the well-known
Bonn potential~\cite{Machleidt:1987hj,Stoks:1994wp}. This
way, the invariance under any rotation in isospin space, is reproduced
accurately. The off shell behaviour of Bonn C-D is based upon the
relativistic Feynman amplitudes for meson-exchange
($\eta, \pi, \rho, \omega, \sigma$ and $\, \phi$ mesons in our case),
a fact that has attractive consequences in nuclear structure
applications~\cite{Machleidt:2000ge}.
%%%

Motivated by its successful application on similar calculations for
various semileptonic nuclear processes
\cite{Kosmas:1992yxv,Kosmas:2001sx,Kosmas:1989pj,Tsakstara:2011zzc}, the QRPA method
is employed in this work to construct explicitly the nuclear ground
state, $\vert gs\rangle \equiv \vert 0^+\rangle$, of the studied
even-even isotope ($^{76}$Ge in our case) through the numerical
solution of the BCS equations. The vector proton (neutron) nuclear
form factors are evaluated as
\begin{equation}
F_{N_n}(Q^2) = \frac{1}{N_n}\sum_j \sqrt{2 j +1}\, \langle j\vert j_0(\vert \mathbf{q}\vert r)\vert j\rangle\left(\upsilon^j_{N_n}\right)^2 \, ,
\end{equation}
where $N_{n}=Z\,\,(\mathrm{or}\,\, N)$ and $\upsilon^j_{N_{n}}$
denotes the occupation probability amplitude of the $j$th
single-nucleon orbit (see e.g. Ref.~\cite{Papoulias:2015vxa}).

%%%%%%%%%%%%%%%%%%%%%%%%%%%%%%%%%%%%%%%%%%%%%%%%%%%%%%%%%%%%%%%%%%%%%%%%%%%%%%%%

\section{Experimental Setups}
\label{sect:experiments}

\subsection{Reactor neutrino experiments}

Recently, it became feasible to detect neutrino-nucleus scattering events by
using high purity germanium-based detectors 
(HPGe detector)~\cite{Soma:2014zgm,Kosmas:2015vsa}. In this work, we are interested in the
possibility of probing the existence of a fourth light sterile neutrino through potential
deviations on the low-energy CE$\nu$NS measurements at reactor neutrino
experimental facilities, such as 
TEXONO~\cite{Kerman:2016jqp,Sevda:2016otj},
$\nu$GeN~\cite{Belov:2015ufh}, GEMMA~\cite{Beda:2013mta},
CONNIE~\cite{Moroni:2014wia,Aguilar-Arevalo:2016qen} and
MINER~\cite{Agnolet:2016zir}. 
We have considered as reference experimental setup 1 kg
of $^{76}$Ge detector and a detection threshold of
100~$\mathrm{eV_{ee}}$
\footnote{$\mathrm{eV_{ee}}$ refers to the electron equivalent
    energy, and should be distinguished from the nuclear recoil
    energy, $\mathrm{eV_{nr}}$ (see
    Sec.~\ref{sect:numer_results}).}.  We note, however, that the
absence of precise information regarding the fuel composition
restricts us to take into account only the dominant component of the
antineutrino spectrum provided by $^{235}$U. In this respect, for the
present study we assume a typical flux of
$\Phi_{\bar{\nu}_{e}} \sim 10^{13}$
$\nu \, \mathrm{ s^{-1} \, cm^{-2}}$ for a detector located at 28 m
from the 2.9 GW reactor core. In order to estimate the emitted
$\bar{\nu}_e$ energy-distribution,
$\eta_{\bar{\nu}_e}^{\mathrm{react}}(E_\nu)$, for energies above 2
MeV, existing experimental data from Ref.~\cite{Mueller:2011nm} are
employed, while for energies $E_{\bar{\nu}_{e}}<2$ MeV existing
theoretical estimations~\cite{Kopeikin:1997ve} are assumed.

\subsection{Spallation Neutron Source experiments}

The Spallation Neutron Source at Oak Ridge~\cite{Scholberg:2005qs} has
been recently considered as a promising facility to measure CE$\nu$NS
events within the SM~\cite{Collar:2014lya,Papoulias:2015vxa} as well
as to explore exotic neutrino
properties~\cite{Papoulias:2013gha,Kosmas:2015sqa,Papoulias:2015iga}. The
COHERENT experiment~\cite{Akimov:2015nza} aims to use intense
neutrino beams (of the order of
$\Phi_{\nu_{\alpha}} \sim 10^{7} \, \nu \, \mathrm{s^{-1} cm^{-2}}$
per flavor) resulting from pion decay. Specifically, the stopped-pion
neutrino beam consists of: (i) monochromatic muon-neutrino $\nu_{\mu}$
flux with energy 29.9 MeV produced via pion decay at rest
$\pi^+ \rightarrow \mu^{+} \nu_{\mu} $ within $\tau=26 \, \mathrm{ns}$
(prompt flux) and (ii) electron neutrinos, $\nu_e$, and muon
antineutrinos, $\bar{\nu}_{\mu}$, that are emitted from the muon-decay
$\mu^{+} \rightarrow \nu_{e} e^{+} \bar{\nu}_{\mu}$ within
$\tau=2.2 \, \mathrm{\mu s}$ (delayed
flux)~\cite{Efremenko:2008an}. The delayed flux is described by the
well-known normalized
distributions~\cite{AguilarArevalo:2008yp,Louis:2009zza}
 \begin{equation}
 \begin{aligned}
\eta_{\nu_{e}}^{\mathrm{SNS}}(E_\nu)=& 96 E_{\nu}^{2}M_{\mu}^{-4} \left( M_{\mu}-2E_{\nu}\right)\, ,\\
\eta_{\bar{\nu}_{\mu}}^{\mathrm{SNS}}(E_\nu)=& 16 E_{\nu}^{2}M_{\mu}^{-4} \left( 3 M_{\mu}-4E_{\nu}\right)\, ,
\end{aligned}
\label{labor-nu}
\end{equation}
with $E_{\nu}^{\text{max}}=M_{\mu}/2$ and $M_{\mu}=105.6 \, \mathrm{MeV}$ denoting the muon rest mass. 

In this work, the calculation is performed for two cases
corresponding to (i) the ``current'' configuration: a ($^{20}$Ne,
$^{40}$Ar, $^{76}$Ge, $^{132}$Xe) target with mass (391, 456, 100,
100)~kg located at (46, 46, 20, 40)~m from the source with energy
threshold of (30, 20, 10, 8)~$\mathrm{keV_{nr}}$ and a running time of
$2.4 \times 10^7$s, and (ii) the ``future'' configuration: 1~ton of
detector mass located at 20~m from the source with energy threshold
$1~\mathrm{keV_{nr}}$ and 1 year of data taking time (see e.g
Ref.~\cite{Kosmas:2015sqa}).

%%%%%%%%%%%%%%%%%%%%%%%%%%%%%%%%%%%%%%%%%%%%%%%%%%%%%%%%%%%%%%%%%%%%%%%%

\section{Neutrino oscillations with a  light sterile neutrino}
\label{sec:impact-light-sterile}
In the present study, we employ a minimal extension of the standard model by considering a
fourth light sterile neutrino state added to the three active
neutrinos. In this case, neutrino flavor eigenstates $\nu_\alpha$,
with $\alpha=\{e,\mu,\tau,s, \cdots \}$ are related to neutrino mass
eigenstates $\nu_i$, with $i= \{1,2,3,4,\cdots \}$ through a unitary
transformation as $\nu_\alpha =\sum_i U_{\alpha i} \nu_i$.
Sterile neutrino mass schemes have been considered in the literature
with various motivations. Attractive possibilities are the early 2+2
models~\cite{peltoniemi:1993ec,peltoniemi:1993ss}. While they still
constitute probably one of the most interesting sterile extensions of
the standard model, their original motivation is gone. On the other
hand, they are strongly restricted by solar and atmospheric data and
do not allow for the eV-scale neutrino mass we are interested in
here~\cite{Maltoni:2003yr,Maltoni:2004ei,Kopp:2011qd}.
  For this reason, we focus on the (3+1) scheme, which does allow for
  eV-neutrino masses as long as the doublet-singlet mixing angles are
  adequately small, so that the sterile state decouples from both
  solar and atmospheric conversions, a possibility absent in the 2+2
  schemes.  
  
The generated reactor antineutrinos $\bar{\nu}_e$ of
energy $E_{\nu}$ are expected to travel the propagation distance $L$
with the survival probability%~\cite{An:2014bik}
\begin{equation}
P_{ee} = 1 - 4 \sum_{i=1}^{3}\sum_{j>i}^{4} \left\vert U_{ei} \right\vert^2 \left\vert U_{ej} \right\vert^2 \sin^2 \left( \Delta_{ji} \right) \, ,
\label{Eq.Pee}
\end{equation}
where $\Delta_{ji} = \Delta m^2_{ji} L /4 E_{\nu}$, with the mass splittings denoted as $\Delta m^2_{ji} = m^2_j - m^2_i$.  
 In this work we will consider values of $\Delta m^2_{ji}$ of the
  order of $1~\mathrm{eV}^2$, as required in order to account for the current
  neutrino anomalies.  The matrix elements entering
Eq.(\ref{Eq.Pee}) take the form%~\cite{deGouvea:2008qk}}
\begin{eqnarray}
U_{e1} & = & \cos \theta_{14} \cos \theta_{13} \cos \theta_{12}\, , \\
U_{e2} & = & \cos \theta_{14} \cos \theta_{13} \sin \theta_{12}\, , \\
U_{e3} & = & \cos \theta_{14} \sin \theta_{13} \, , \\
U_{e4} & = & \sin \theta_{14} \, .
\end{eqnarray}
In this framework, the hypothesis of a fourth neutrino generation
yields the approximate electron neutrino survival probability  for a given value of ($L/E_\nu$)
\begin{equation}
\begin{aligned}
 P_{ee} \simeq 1 -& \cos^4 \theta_{14} \sin^2 2 \theta_{13} \sin^2 \left( \frac{\Delta m_{31}^2 L}{4 E_{\nu}} \right) \\
  -& \sin^2  2 \theta_{14} \sin^2 \left( \frac{\Delta m_{41}^2 L}{4 E_{\nu}} \right) \, .
\end{aligned}
\end{equation}
Note that, for vanishing $\theta_{14}$ or neutrino paths larger than 100 m, the latter expression reduces
to the well-known oscillation probability for short-baselines probed at the new generation 
of reactor experiments such as Daya Bay~\cite{An:2013zwz}, RENO~\cite{Ahn:2012nd} and Double Chooz~\cite{Abe:2011fz}.
On the contrary, at shorter distances, atmospheric neutrino driven oscillations can be neglected and
 the neutrino survival probability can be effectively parametrized as
\begin{equation}
 P_{ee} = 1 - \sin^2  2 \theta_{14} \sin^2 \left( \frac{\Delta m_{41}^2 L}{4 E_{\nu}} \right) \, .
\end{equation}
 %

%
%%%%%%%%%%%%%%%%%%%%%%%%%%%%%%%%%%%%%%%%%%%%%%%%%%%%%%%%%%%%%%%%%%%%%%%
\begin{figure}[t]
\centering
\includegraphics[width= \linewidth]{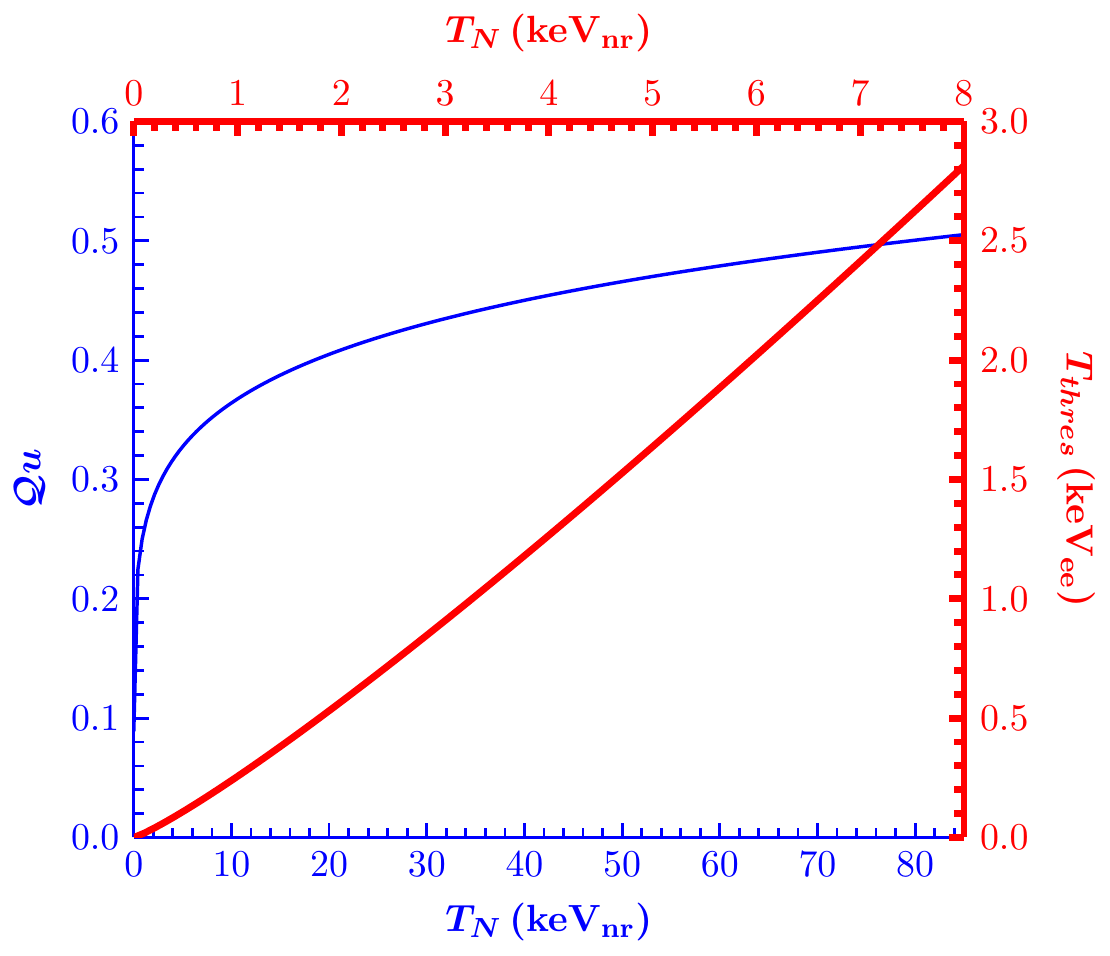}
\caption{(\textit{Blue labeling}): The quenching factor,
  $\mathcal{Q}u(T_N)$ for $^{76}$Ge and (\textit{Red labeling}): the
  equivalent electron energy as a function of the nuclear recoil
  energy, $T_N$.}
\label{fig:quenching-factor}
\end{figure}
%%%%%%%%%%%%%%%%%%%%%%%%%%%%%%%%%%%%%%%%%%%%%%%%%%%%%%%%%%%%%%%%%%%%%%%%%%%%
%
%%%%%%%%%%%%%%%%%%%%%%%%%%%%%%%%%%%%%%%%%%%%%%%%%%%%%%%%%%

\section{Numerical results}
\label{sect:numer_results} 

Reactor neutrino experiments are sensitive to the mixing matrix element $U_{e4}$, while SNS experiments are sensitive to both $U_{e4}$ and $U_{\mu 4}$,
through the measurement of $\sin^2 2 \theta_{14}$. 
In the presence of sterile neutrinos, the differential event rate in
terms of the nuclear recoil energy $T_N$, reads
\begin{equation}
\begin{aligned}
\frac{dN_{\mathrm{sterile}}^{events}}{dT_N}=& K \int_{E_{\nu_\mathrm{min}}}^{E_{\nu_\mathrm{max}}} dE_{\nu} \, \eta^{\lambda}_{\nu_{\alpha}}(E_\nu) P_{\alpha \alpha}(E_\nu) \\
& \times \int_{-1}^{1}   d\cos \theta \,   \frac{d\sigma_{\nu_\alpha}} {d\cos\theta} 
\delta\left(T_N - \frac{Q^2}{2 M}\right) \,,\\  \quad \lambda=\mathrm{react, \, SNS}\, ,
\end{aligned}
\end{equation}
where $M$ is the nuclear mass and $K = N_{targ} \Phi_{\nu_{\alpha}} t_{tot} $, with $N_{targ}$
denoting the total number of atoms in the detector and
$t_{tot}$ the time window of exposure. 
The incident neutrino flux is given by $\Phi_{\nu_{\alpha}}$,
while $\eta^{\mathrm{react}}_{\nu_{\alpha}}$ and
$\eta^{\mathrm{SNS}}_{\nu_{\alpha}}$ denote the neutrino energy-distributions at reactor experiments and SNS, respectively. Note that, in contrast to our previous studies~\cite{Kosmas:2015sqa,Kosmas:2015vsa}, the above expression includes the effect of flavor oscillations in the neutrino propagation. Then, the number of events for a given detector
threshold, $T_{\mathrm{thres}}$, is evaluated through the integral
\begin{equation}
N_{\mathrm{sterile}}^{events} = \int_{T_{\mathrm{thres}}}^{T_\mathrm{max}}
\frac{dN_{\mathrm{sterile}}^{events}}{dT_N}\, dT_N\, ,
\end{equation}
where $T_\mathrm{max}$ is the maximum recoil energy obtained from the
kinematics of the process~\cite{Papoulias:2013gha}.

%
%%%%%%%%%%%%%%%%%%%%%%%%%%%%%%%%%%%%%%%%%%%%%%%%%%%%%%%%%%%%%%%%%%%%%%%%%%%%
\begin{figure}[t]
\centering
\includegraphics[width= \linewidth]{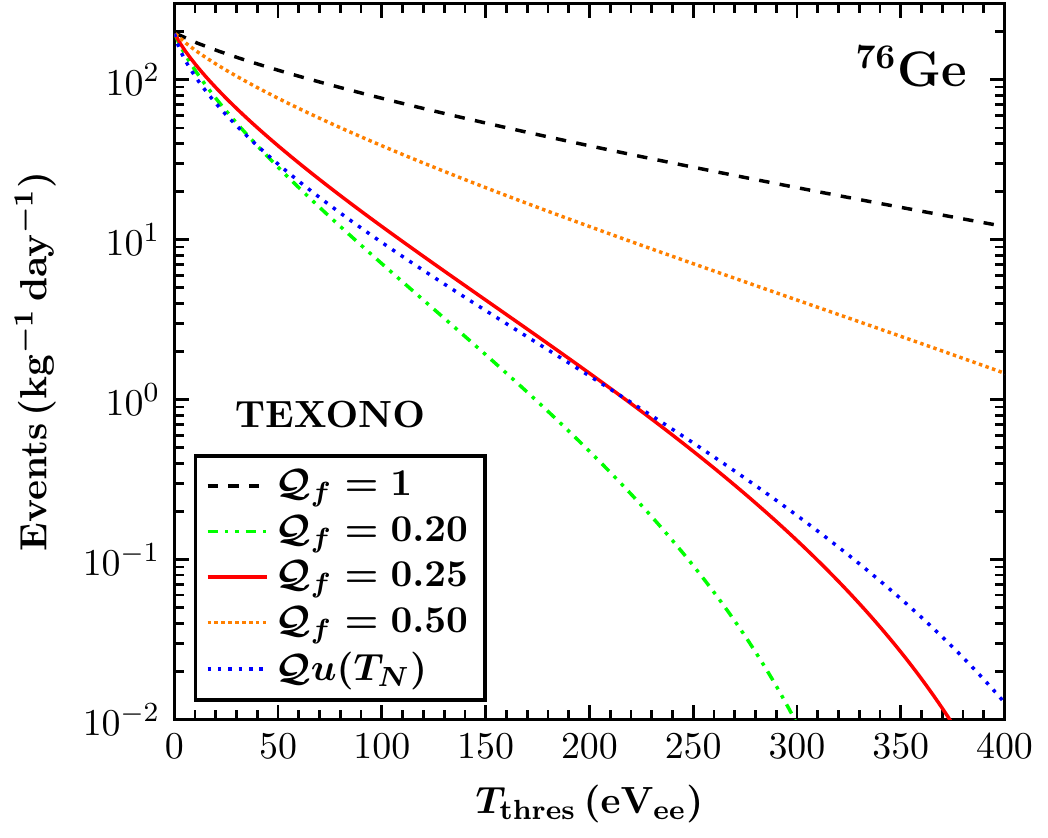}
\includegraphics[width= \linewidth]{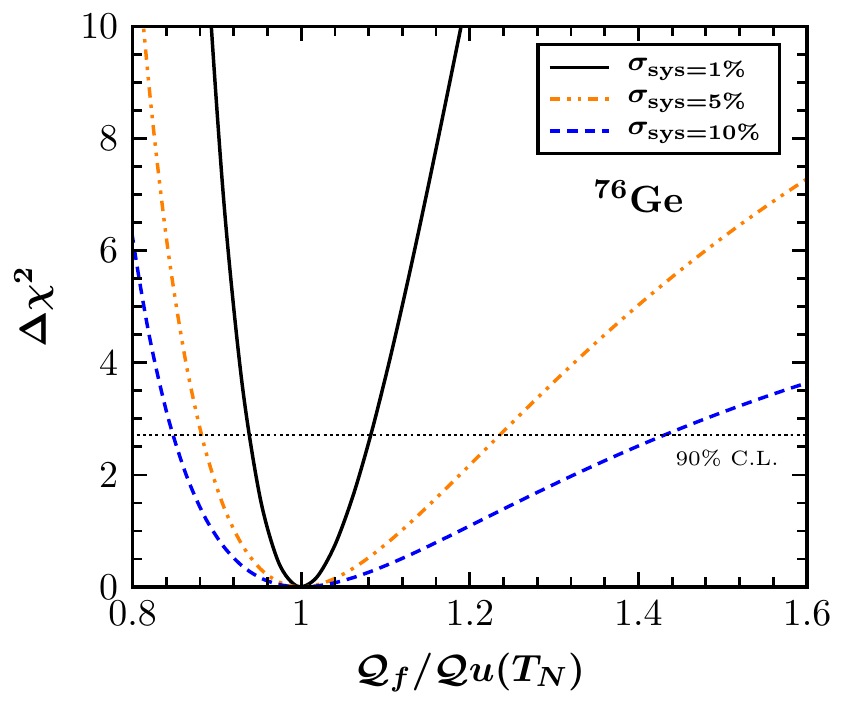}
\caption{Top panel: CE$\nu$NS events within the SM as a function of
  the detector threshold assuming different quenching factors and a
  1kg-day $^{76}$Ge target. A notable agreement is verified between
  the results obtained for the case of constant quenching factor in
  the range $\mathcal{Q}_f=0.20-0.25$ and the empirical quenching
  factor of Eq.(\ref{eq:Qu}). Bottom panel: Sensitivity of the TEXONO
  experiment to the quenching factor $\mathcal{Q}_f$ normalized to the
  empirical quenching factor of Eq.(\ref{eq:Qu}) for various systematic
  errors and a background of 1~cpd (see the text).}
\label{fig:TEXONO-quenched-events}
\end{figure}
%%%%%%%%%%%%%%%%%%%%%%%%%%%%%%%%%%%%%%%%%%%%%%%%%%%%%%%%%%%%%%%%%%%%%%%%%%%%
%

%%%%%%%%%%%%%%%%%%%%%%%%%%%%%%%%%%%%%%%%%%%%%%%%%%%%%%%%%%%%%%%%%%%%%%%%%%%%
\begin{figure}[t!]
\centering
\includegraphics[width=  \linewidth]{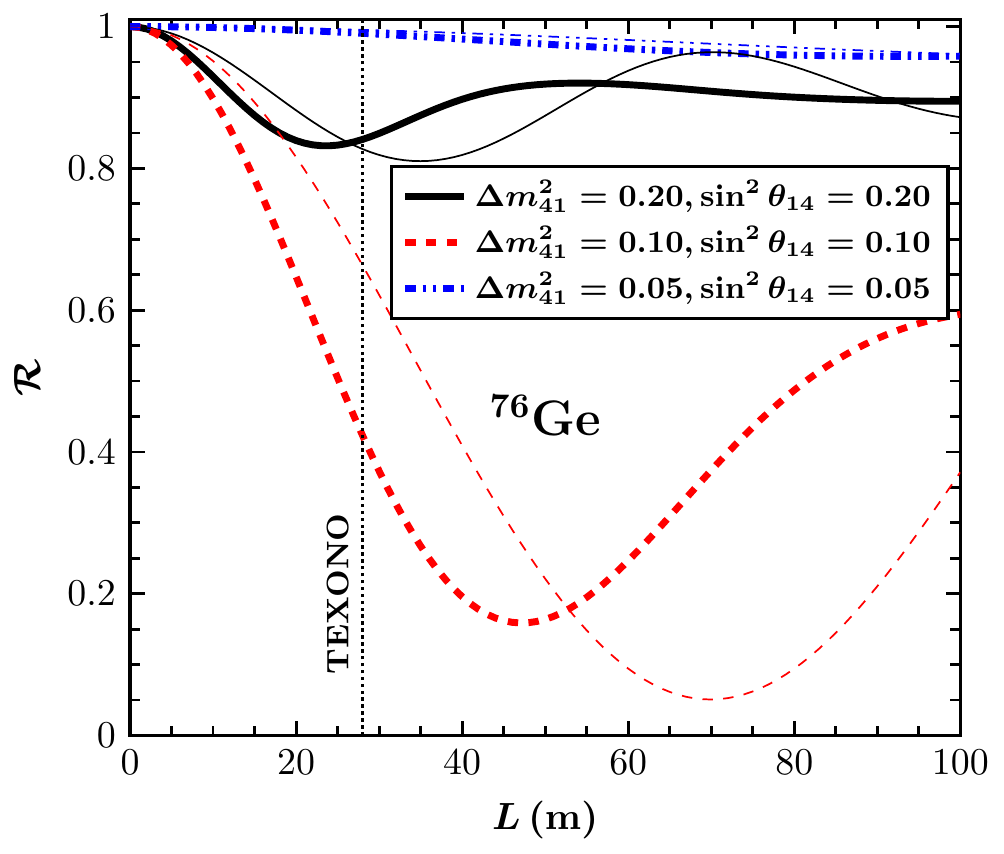}
\caption{Ratio
    $\mathcal{R} =
    N_{\mathrm{sterile}}^{events}/N_{\mathrm{SM}}^{events}$
    for a detector threshold $T_{\mathrm{thres}}=100~\mathrm{eV_{ee}}$
    as a function of the baseline $L$, at the TEXONO experiment.  The quenching effect is
    considered (neglected) in the thin (thick) lines.  The vertical
  dotted line indicates the TEXONO baseline.}
\label{fig.ratio}
\end{figure}
%%%%%%%%%%%%%%%%%%%%%%%%%%%%%%%%%%%%%%%%%%%%%%%%%%%%%%%%%%%%%%%%%%%%%%%%%%%%
%%
%%%%%%%%%%%%%%%%%%%%%%%%%%%%%%%%%%%%%%%%%%%%%%%%%%%%%%%%%%%%%%%%%%%%%%%%%%%%
\begin{figure*}[t!]
\centering
\includegraphics[width= 0.48\linewidth]{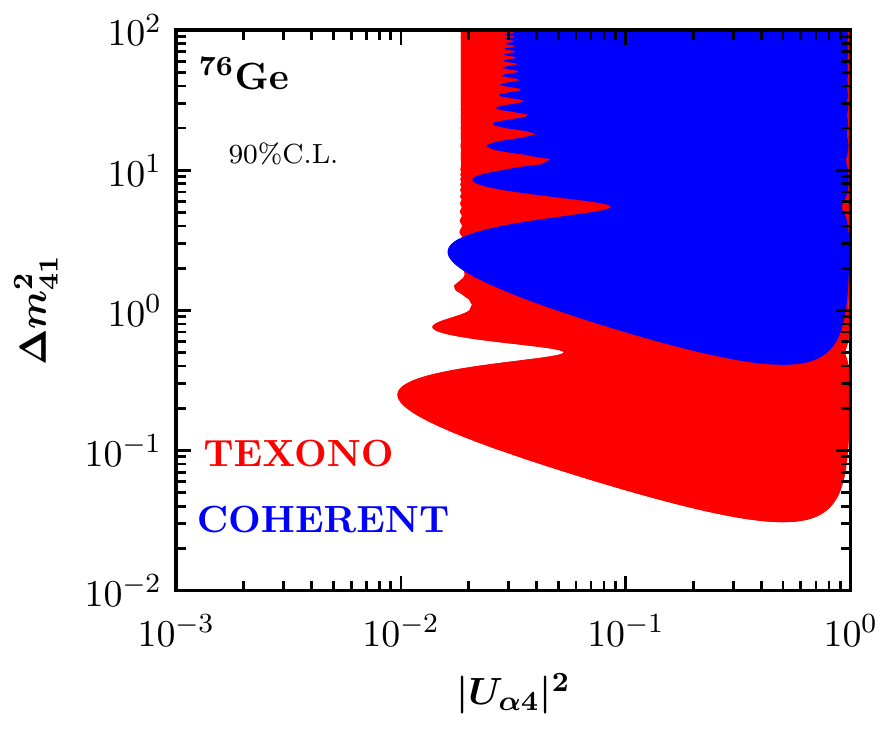}
\includegraphics[width= 0.48\linewidth]{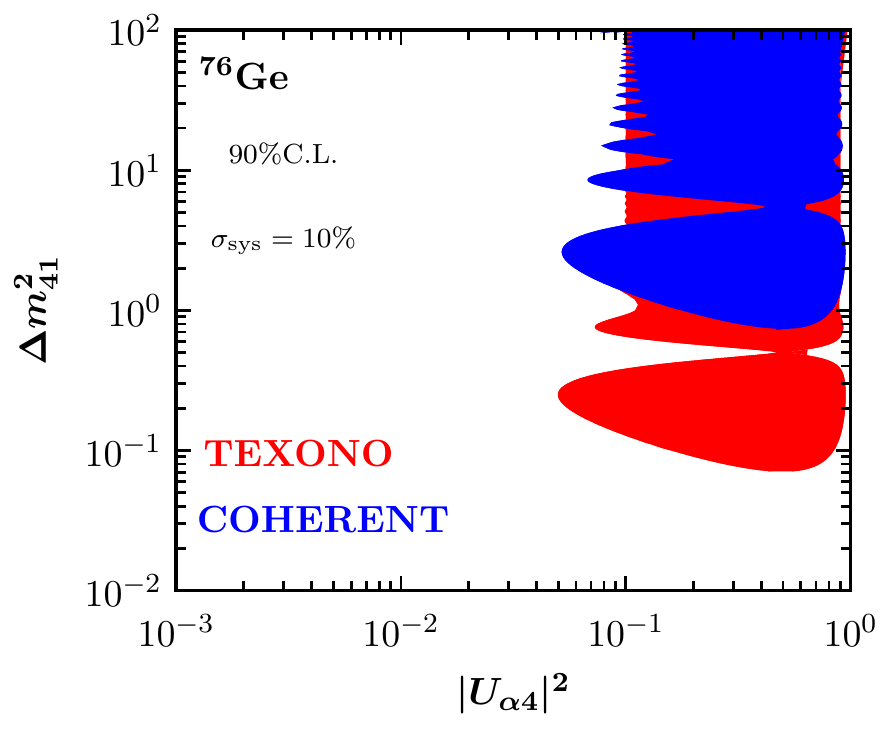}
\caption{90\% C.L.  sensitivity regions in the
  $(\vert U_{\alpha 4}\vert^2,\, \Delta m^2_{41})$ planes with $\alpha=e$ (red region) and $\alpha=\mu$ (blue region) assuming a light
  sterile neutrino in the (3+1) scheme, at the TEXONO  and
  COHERENT experiments respectively (for details see the
  text). In the left panel systematic uncertainties and background events are neglected, while in the right panel the calculation assumes a systematic error of $\sigma_{\mathrm{sys}}=10\%$ for the corresponding background events in each experiment.}
\label{fig.Ua4}
\end{figure*}
%%%%%%%%%%%%%%%%%%%%%%%%%%%%%%%%%%%%%%%%%%%%%%%%%%%%%%%%%%%%%%%%%%%%%%%%%%%%
%

Focusing on the relevant CE$\nu$NS experiments, the detectable energy is
lower than the energy imparted to the nuclear target ($\mathrm{eV_{nr}}$),
since the employed  detectors are sensitive to an ionization
energy equivalent to an electron energy
($\mathrm{eV_{ee}}$)~\cite{Giomataris:2005fx}. To account for the
energy loss due to the conversion to phonons in such measurements, the
present calculations take into consideration the quenching effect on
the nuclear recoil events by multiplying the energy scale by a
quenching factor, $\mathcal{Q}_f$~\cite{Vergados:2009ei}. In general,
$\mathcal{Q}_f$ varies with the nuclear recoil
energy and, usually, for its estimation the
following empirical form is considered~\cite{Simon:2002cw}:
\begin{equation}
\mathcal{Q}u(T_N) = r_1 \left[ \frac{T_N}{1 \mathrm{keV}}\right]^{r_2}, \quad r_1 \simeq 0.256, \, \, r_2 \simeq 0.153 \, .
\label{eq:Qu}
\end{equation}
The dependence of $\mathcal{Q}_f$ on the nuclear recoil energy, $T_N$,
is shown in Fig.~\ref{fig:quenching-factor} where the equivalent
electron energy as a function of $T_N$ is also presented. The top
panel of Fig.~\ref{fig:TEXONO-quenched-events}, illustrates the
variation of the expected CE$\nu$NS event rates at different
thresholds and quenching factors at the TEXONO reactor experiment. On
the other hand, in the bottom panel of
Fig.~\ref{fig:TEXONO-quenched-events}, based on the
$\chi^2(\mathcal{Q}_f) = \min_\xi \left[\chi^2 (\mathcal{Q},
  \xi)\right]$
function we examine how well the quenching factor is required to be
known in order to record a clear signal, assuming SM interactions
only. For this calculation, a 1~kg $^{76}$Ge detector has been 
assumed with a threshold of 100~$\mathrm{eV_{ee}}$, one year of
exposure and various systematic errors. Following
Ref.~\cite{Soma:2014zgm}, the considered background is set at 1~cpd e.g. 1 event~
$\mathrm{day^{-1}~ kg^{-1}~keV^{-1}}$ ensuring a signal-to-noise ratio
$>22$ (for its spectral shape the reader is referred to
Ref~\cite{Lindner:2016wff}).

%
%%%%%%%%%%%%%%%%%%%%%%%%%%%%%%%%%%%%%%%%%%%%%%%%%%%%%%%%%%%%%%%%%%%%%%%%%%%%
\begin{figure*}[t]
\centering
\includegraphics[width= 0.8\linewidth]{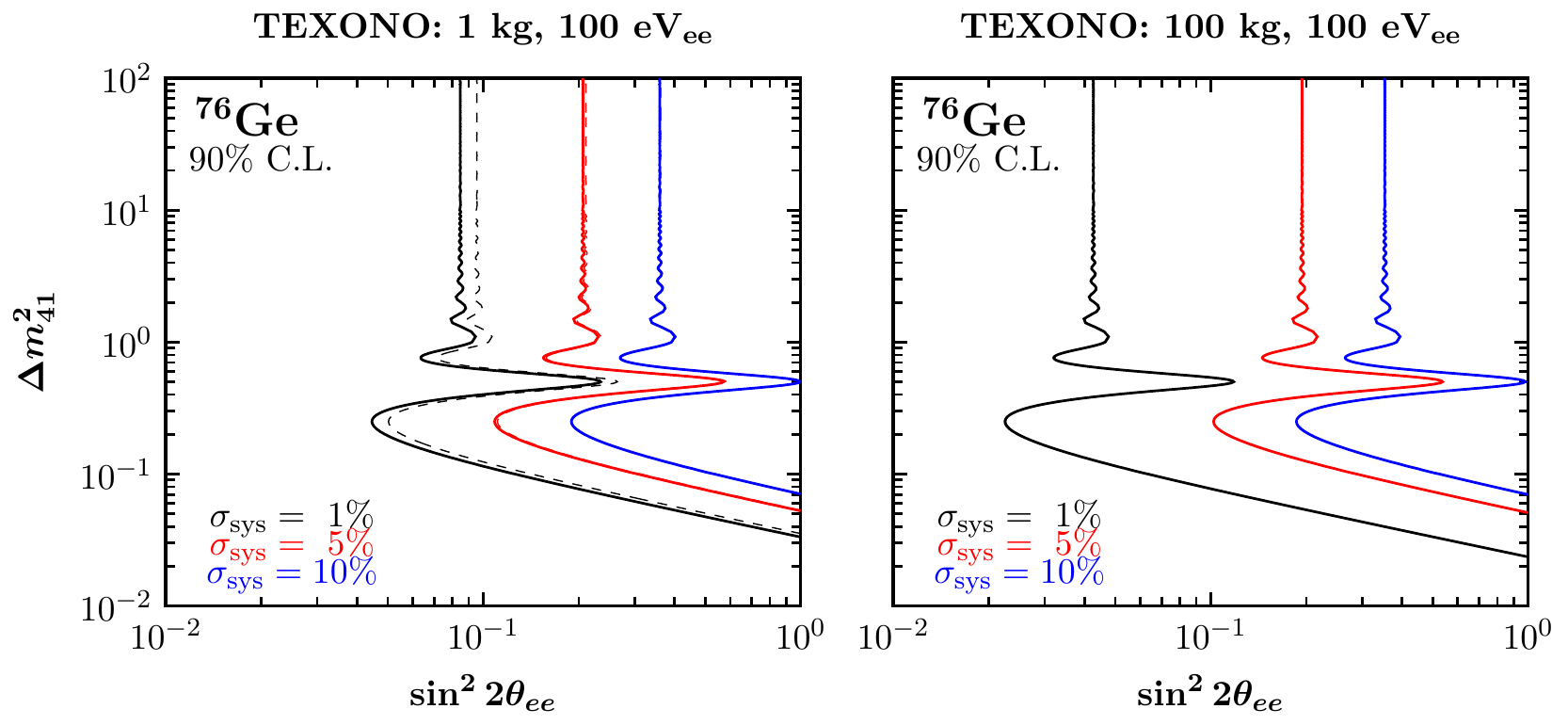}\\
\includegraphics[width= 0.8\linewidth]{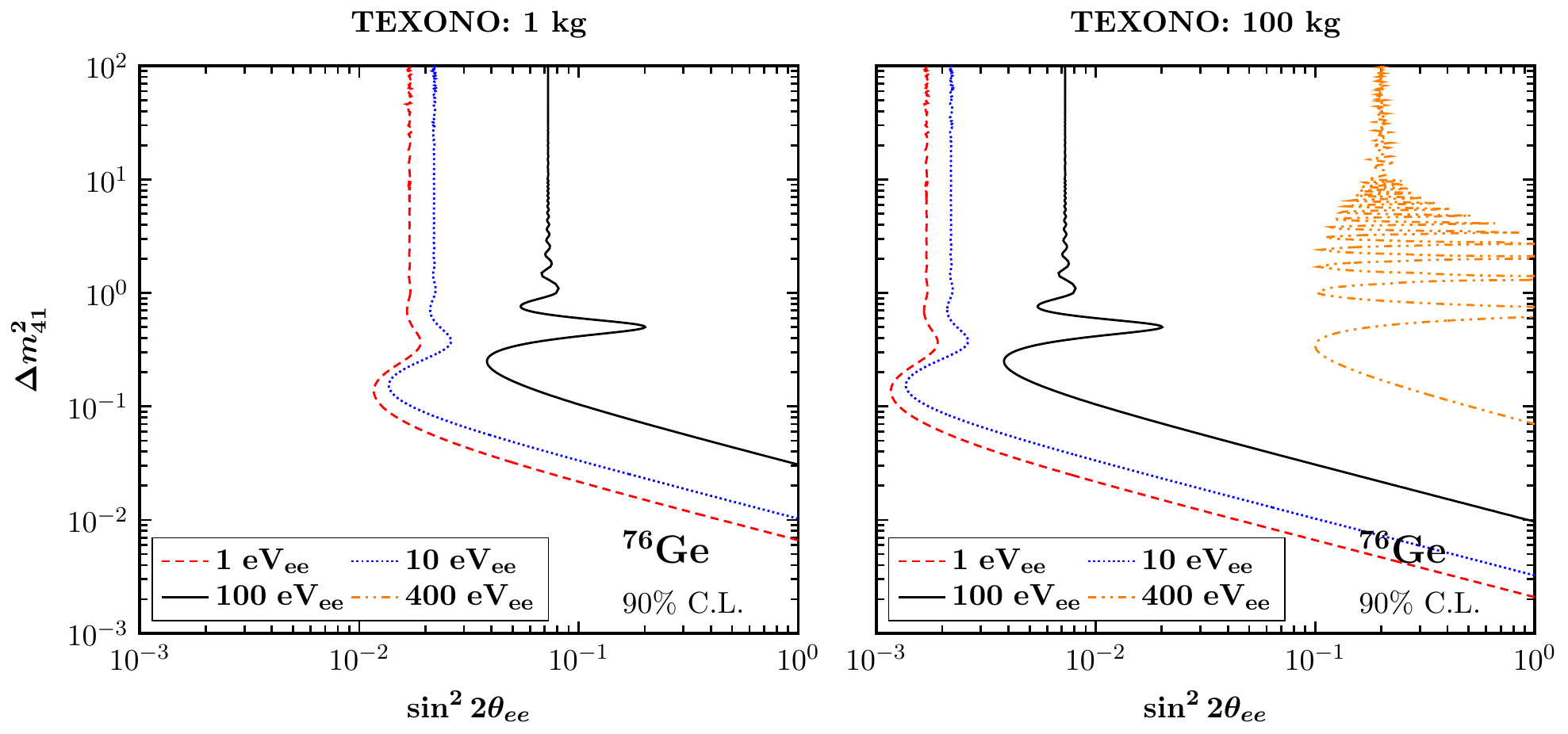}
\caption{Top panel: 90\% C.L. sensitivity region in the
  $(\sin^2 2 \theta_{ee}, \, \Delta m_{41}^2)$ plane for a light
  sterile neutrino in the (3+1) scheme, considering the current
  experimental setup at the TEXONO experiment for different values of
  systematic error for a background level of 1~cpd (solid lines)
    and 10~cpd (dashed lines). Bottom panel: 90\% C.L. sensitivity
  region for different configurations of detector mass and operation
  threshold with neglected background and systematic uncertainties.}
\label{fig.sterile.sys}
\end{figure*}
%%%%%%%%%%%%%%%%%%%%%%%%%%%%%%%%%%%%%%%%%%%%%%%%%%%%%%%%%%%%%%%%%%%%%%%%%%%%
%

In order to get an idea of how the presence of sterile neutrinos
affects the expected number of events at a given detector, we define
the ratio
\begin{equation}
\mathcal{R} = \frac{N_{\mathrm{sterile}}^{events}}{N_{\mathrm{SM}}^{events}} \, ,
\end{equation}
i.e., the portion of events originated from sterile neutrinos in the
total number of SM events.  We mention that $\mathcal{R}$ is independent of the
detector mass and may also limit inevitable flux uncertainties. Apparently, the
equality
$\mathcal{R} = \vev{
  \sigma}_{\mathrm{sterile}}/\vev{\sigma}_{\mathrm{SM}}$
holds true, where $\vev{ \sigma}_{\mathrm{SM}}$ stands for the
  SM cross section averaged over the reactor neutrino flux
  distribution, while $\vev{ \sigma }_{\mathrm{sterile}}$ 
  denotes the corresponding flux-averaged cross section that includes
  also the oscillation probability.
Figure~\ref{fig.ratio} shows the variation of $\mathcal{R}$ with the distance $L$ for
various choices of the sterile neutrino parameters, assuming a
  $^{76}$Ge detector with mass 1~kg and an energy threshold of
  $T_{\mathrm{thres}}=100~\mathrm{eV_{ee}}$ at the TEXONO experiment. The quenching effect is
  taken into account, while for comparison, the corresponding results
  obtained by neglecting the quenching effect are also illustrated.

The relevant experiments searching for CE$\nu$NS are subject to a number of uncertainties that should be effectively taken into account in order to come out with realistic estimates of the sensitivity to possible new physics phenomena. In such type of experiments the dominant contributions to systematic uncertainties are linked to the lack of precise knowledge on the neutrino flux, the quenching factor, the detector threshold, mass and performance, distance from the source, etc~\cite{Lindner:2016wff}. Background uncertainties depend on the various experimental setups and include mostly beam-related backgrounds (e.g. neutrino-induced neutrons), internal beta- and gamma-radioactivity and other secondary backgrounds from shielding materials~\cite{Dent:2016wcr}. Based on the pull method, in our attempt to quantify the sensitivity of a given CE$\nu$NS experiment
to sterile neutrinos, we define the $\chi^2$ function
\begin{equation}
\begin{aligned}
& \chi^2(\sin^2 2\theta, \Delta m^2_{41}) =   \min_\xi \left[\chi^2 (\sin^2 2\theta, \Delta m^2_{41}, \xi)\right]  \\ &= \min_\xi \left[ \left( \frac{N_{\mathrm{SM}}^{events} - N_{\mathrm{sterile}}^{events} (1+\xi)}{\sigma_{\mathrm{stat}}}\right)^2 + \left( \frac{\xi}{\sigma_{\mathrm{sys}}}\right)^2 \right] \, ,
\end{aligned}
\end{equation} 
with
$\sigma_{\mathrm{stat}} = \sqrt{N_{\mathrm{SM}}^{events} +
  N_{\mathrm{bkg}}^{events}}$,
minimized over the nuisance parameter $\xi$. Following a conservative
approach, in this work we adopt typical values to account for the
systematic error, i.e. $\sigma_{\mathrm{sys}}=10\%$.  For the case of TEXONO we consider a
background level of 1~cpd, while for the case of COHERENT we assume
that the number of background events is of the order of $20\%$ of the
$N_{\mathrm{SM}}^{events}$~\cite{Akimov:2015nza}. For convenience it
is also useful to obtain the minimum $\chi^2$ with respect to $\xi$
analytically, as
\begin{equation}
\chi^2_{\mathrm{min}} (\sin^2 2\theta, \Delta m^2_{41})  =  \frac{\left(N_{\mathrm{SM}}^{events} - N_{\mathrm{sterile}}^{events}\right)^2}{\sigma_{\mathrm{stat}}^2 + \left( \sigma_{\mathrm{sys}} \, N_{\mathrm{sterile}}^{events}\right)^2}\, .
\end{equation}
Note that, by neglecting the systematic uncertainty, the latter reduces to the simple $\chi^2$ form employed in our previous works~\cite{Kosmas:2015vsa,Kosmas:2015sqa}.

We mention that, due to the smallness of $\theta_{13}$, recently
measured at Daya Bay~\cite{An:2013zwz}, for simplicity in our
calculations we set $\sin^2 2\theta_{13} = 0$. Moreover, we use the
fact that, within the framework of the (3+1) scheme, it holds
\begin{eqnarray}
\sin^2 2 \theta_{\alpha \alpha} &=& 4 \vert U_{\alpha 4} \vert^2 \left( 1 - \vert U_{\alpha 4}\vert^2 \right) \, , \\
\sin^2 2 \theta_{\alpha \beta} &=& 4 \vert U_{\alpha 4} \vert^2 \vert U_{\beta 4} \vert^2 \, ,
\label{sin_ab}
\end{eqnarray}
where $\alpha, \beta = e, \mu, \tau, s$.  Focusing on the relevant
short-baseline (SBL) neutrino experiments, the above expressions enter
into the respective effective survival and transition probabilities,
valid for neutrinos and antineutrinos
\begin{equation}
\begin{aligned}
P_{   {\alpha \alpha}   } &= 1 - \sin^2 2 \theta_{\alpha \alpha} \sin^2 \left(\frac{\Delta m^2_{41} L}{4 E} \right) \, , \\
P_{   {\alpha \beta}   } &=  \sin^2 2 \theta_{\alpha \beta} \sin^2 \left(\frac{\Delta m^2_{41} L}{4 E} \right).
\end{aligned}
\label{P_nu_alpha}
\end{equation}
In Fig.~\ref{fig.Ua4} we illustrate the  90\%
  C.L. sensitivity contours in the $(\vert U_{e4}\vert^2,\, \Delta m^2_{41})$
  plane for the TEXONO experiment, obtained from a two-parameter
  $\chi^2$  analysis as described above and by taking into
  account the quenching effect. The present calculations consider a
  $^{76}$Ge detector with: 1~kg mass, $100~\mathrm{eV_{ee}}$ energy
  threshold and one year of data collection time. For comparison, also
  shown is the corresponding sensitivity region in the
  $(\vert U_{\mu4}\vert^2,\, \Delta m^2_{41})$ plane for the case of
  the COHERENT experiment assuming its ``current'' setup (see Sec.~\ref{sect:experiments}).

%
% 
%%%%%%%%%%%%%%%%%%%%%%%%%%%%%%%%%%%%%%%%%%%%%%%%%%%%%%%%%%%%%%%%%%%%%%%%%%%%
\begin{figure*}[t]
\centering
\includegraphics[width= 0.9\linewidth]{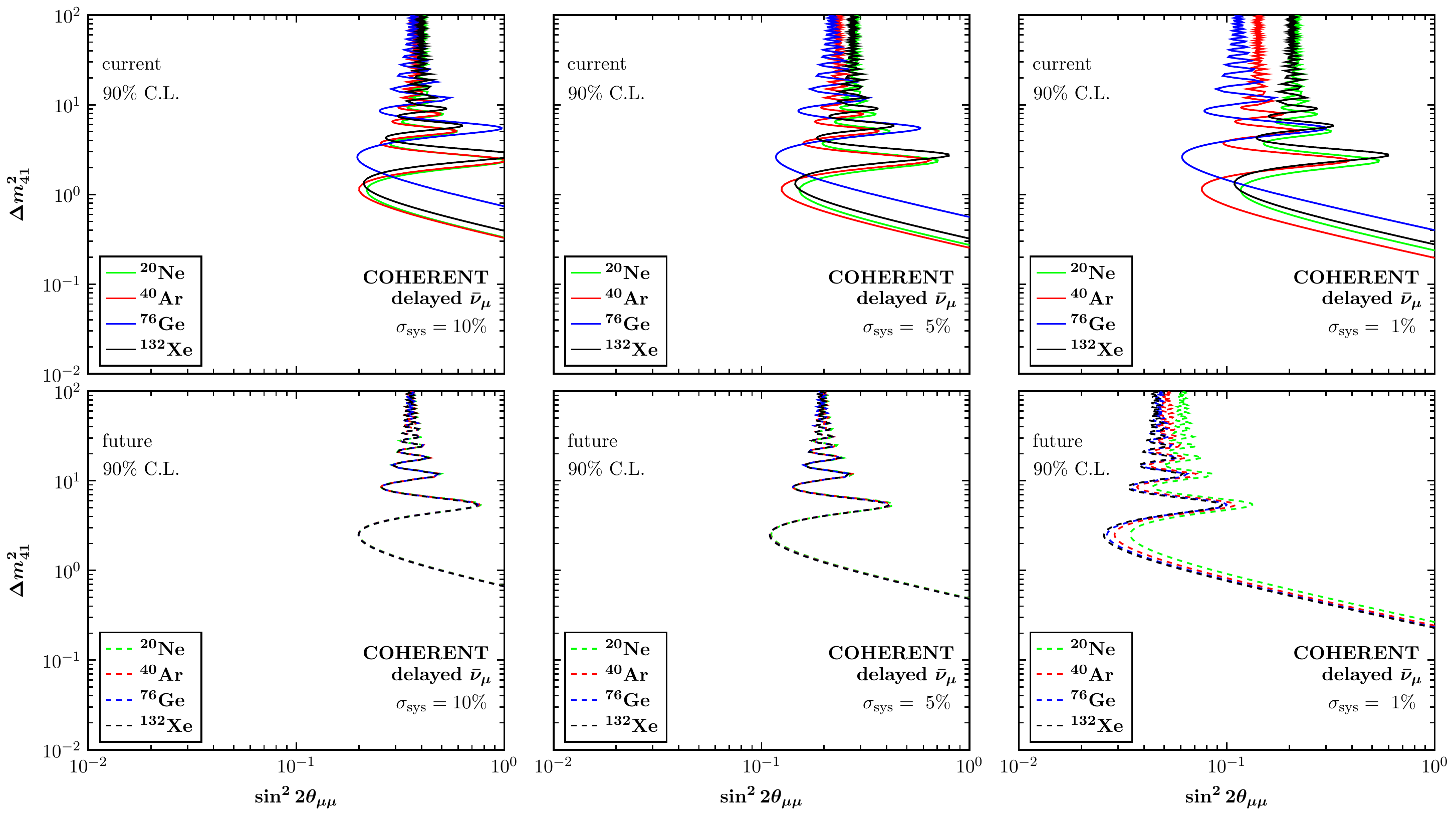}\\
\includegraphics[width= 0.33 \linewidth]{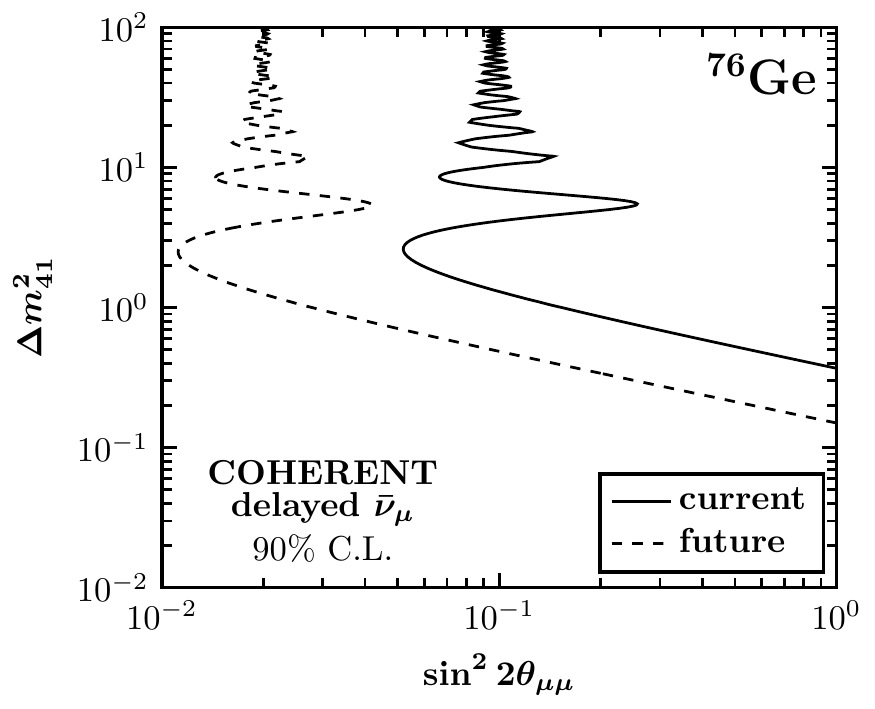}
\caption{Top panel: 90\% C.L. sensitivity region in the
  $(\sin^2 2 \theta_{\mu \mu }, \, \Delta m_{41}^2)$ plane assuming a
  light sterile neutrino in the (3+1) scheme at the COHERENT
  experiment, for various nuclear targets. Different systematic uncertainties are considered while the assumed background events are 20\% of the SM events.  Only the delayed $\bar{\nu}_\mu$ beam is taken into
  account for the ``current'' and ``future'' experimental setup. Bottom panel: Same as above, but with zero systematic error and  neglected backgrounds. The calculation refers to the case of a $^{76}$Ge target only.}
\label{fig.sterile-COHERENT.sys}
\end{figure*}
%%%%%%%%%%%%%%%%%%%%%%%%%%%%%%%%%%%%%%%%%%%%%%%%%%%%%%%%%%%%%%%%%%%%%%%%%%%%
%
%
%%%%%%%%%%%%%%%%%%%%%%%%%%%%%%%%%%%%%%%%%%%%%%%%%%%%%%%%%%%%%%%%%%%%%%%%%%%%
%\begin{figure}[t]
%\centering
%\includegraphics[width= \linewidth]{sterile-COHERENT-new.pdf}
%\caption{90\% C.L. sensitivity region in the
%  $(\sin^2 2 \theta_{\mu \mu }, \, \Delta m_{41}^2)$ plane assuming a
%  light sterile neutrino in the (3+1) scheme at the COHERENT
%  experiment.  Only the delayed $\bar{\nu}_\mu$ beam is taken into
%  account for the ``current'' and ``future'' experimental setup.}
%\label{fig.sterile-COHERENT}
%\end{figure}
%%%%%%%%%%%%%%%%%%%%%%%%%%%%%%%%%%%%%%%%%%%%%%%%%%%%%%%%%%%%%%%%%%%%%%%%%%%%
%
%
%%%%%%%%%%%%%%%%%%%%%%%%%%%%%%%%%%%%%%%%%%%%%%%%%%%%%%%%%%%%%%%%%%%%%%%%%%%%
\begin{figure*}[t]
\centering
\includegraphics[width= 0.40\linewidth]{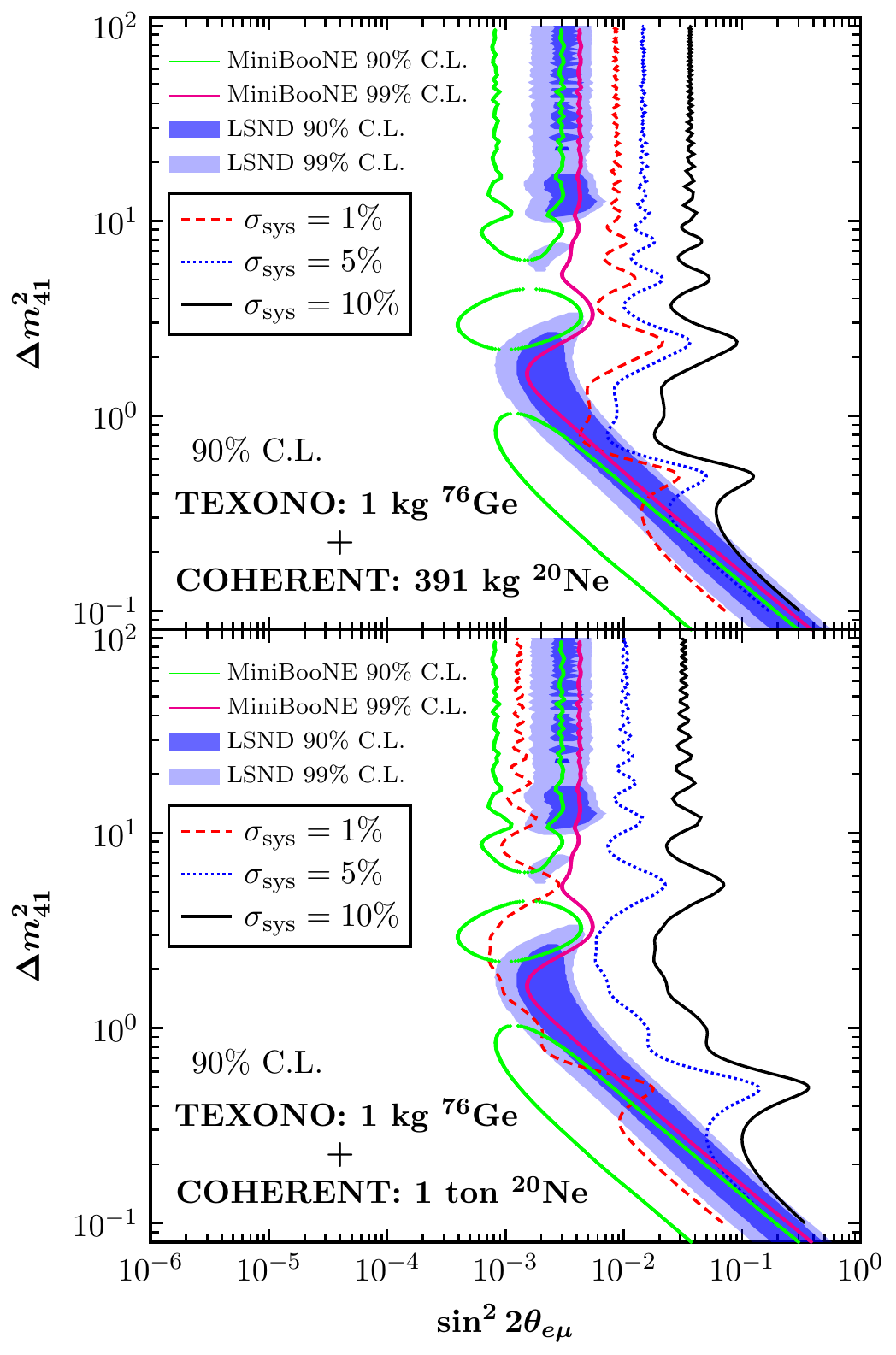}
\includegraphics[width= 0.40\linewidth]{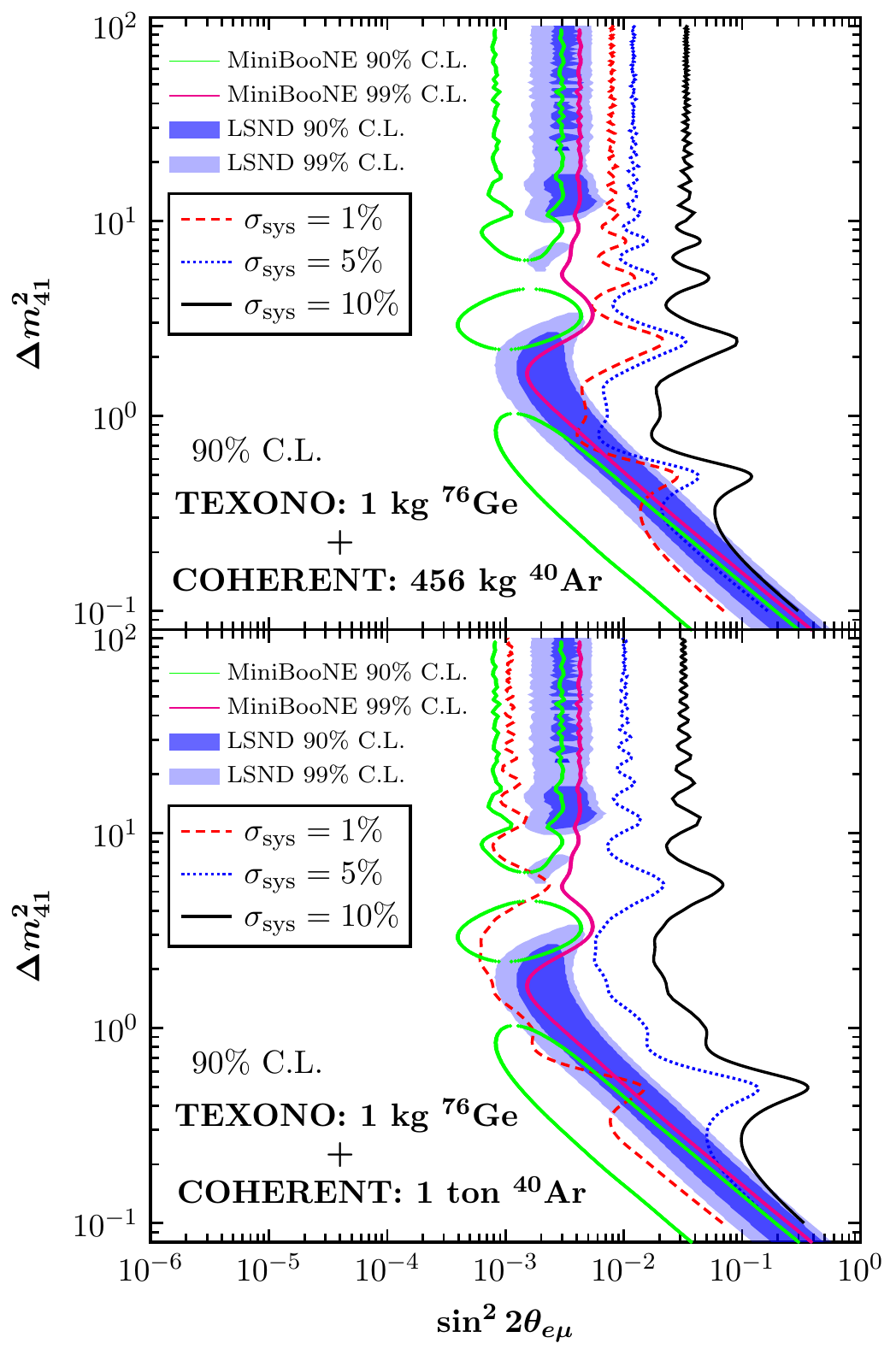}
\includegraphics[width= 0.40\linewidth]{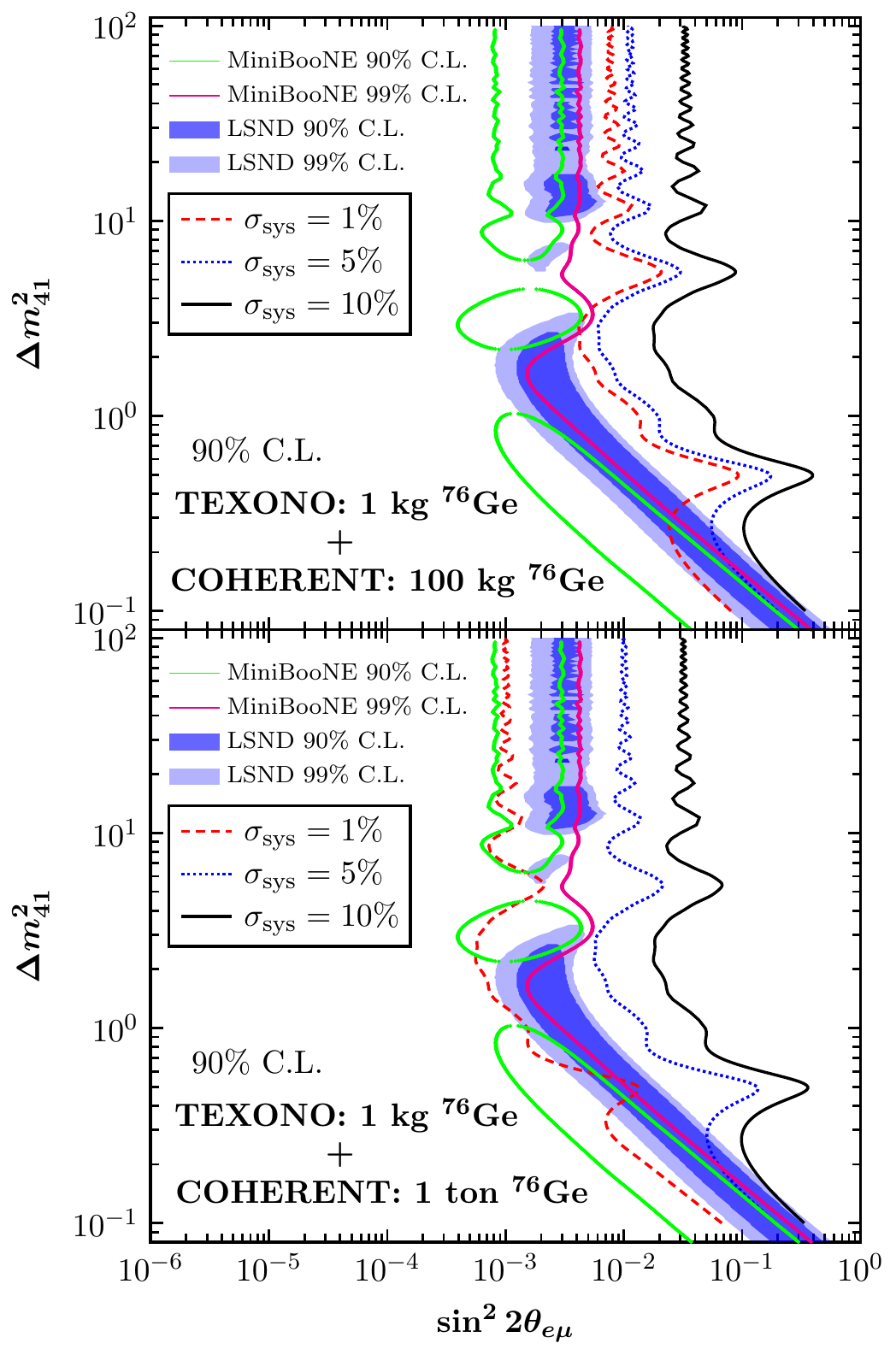}
\includegraphics[width= 0.40\linewidth]{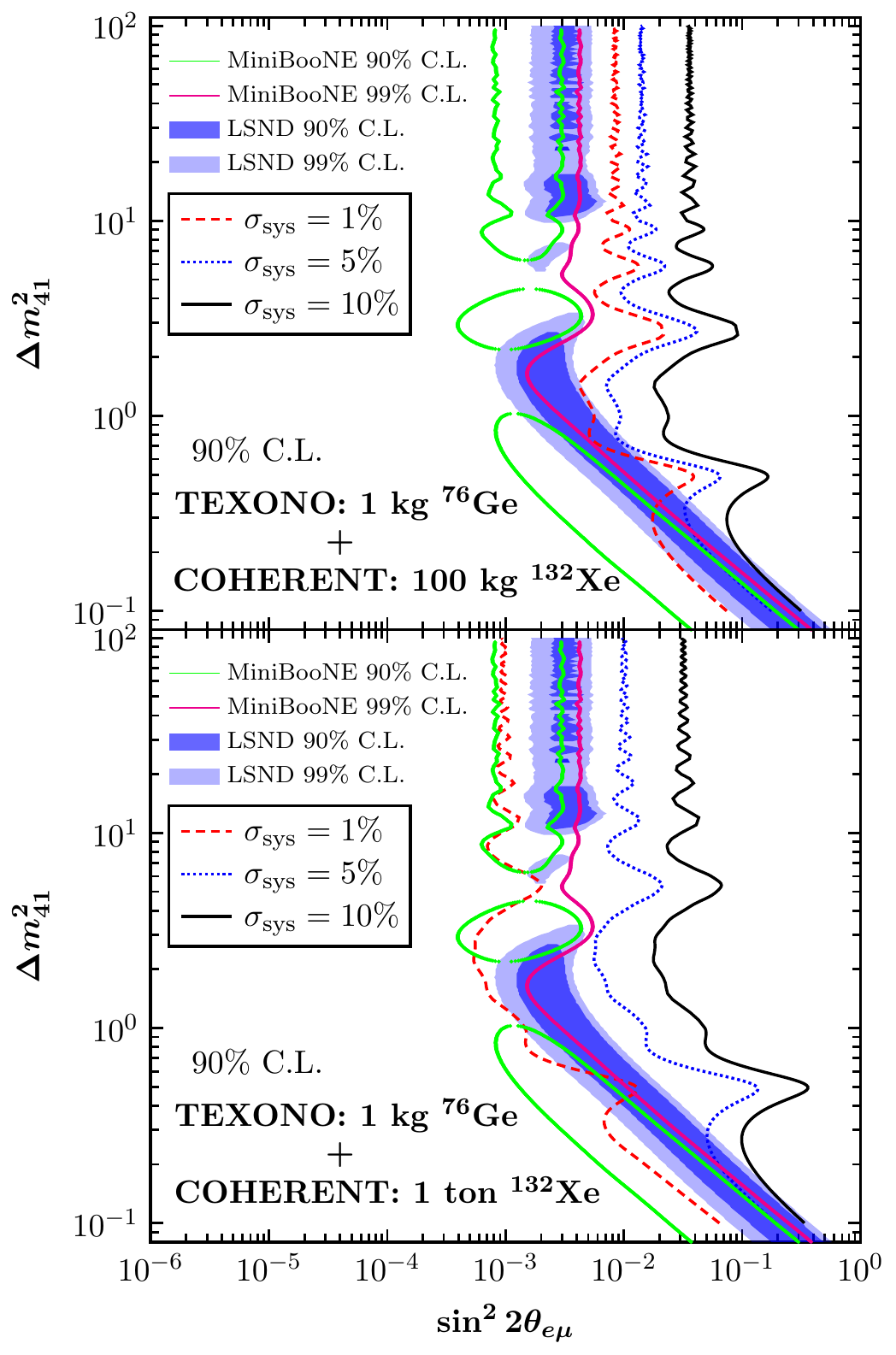}
\caption{90\% C.L sensitivity regions in the
  $(\sin^2 2 \theta_{e \mu }, \, \Delta m_{41}^2)$ plane from a
  combined analysis of COHERENT and TEXONO in the (3+1)
  scheme. Different experimental setups for the COHERENT experiment
  have been considered incorporating systematic uncertainties and
  backgrounds. For comparison the latest allowed regions from the
  LSND~\cite{Athanassopoulos:1995iw,Aguilar:2001ty} and
  MiniBooNE~\cite{Aguilar-Arevalo:2012fmn,Aguilar-Arevalo:2013pmq}
  experiments are also shown.}
\label{fig.theta_emu.sys}
\end{figure*}
%%%%%%%%%%%%%%%%%%%%%%%%%%%%%%%%%%%%%%%%%%%%%%%%%%%%%%%%%%%%%%%%%%%%%%%%%%%%
%
%
%%%%%%%%%%%%%%%%%%%%%%%%%%%%%%%%%%%%%%%%%%%%%%%%%%%%%%%%%%%%%%%%%%%%%%%%%%%%
\begin{figure*}[t]
\centering
\includegraphics[width= 0.8 \linewidth]{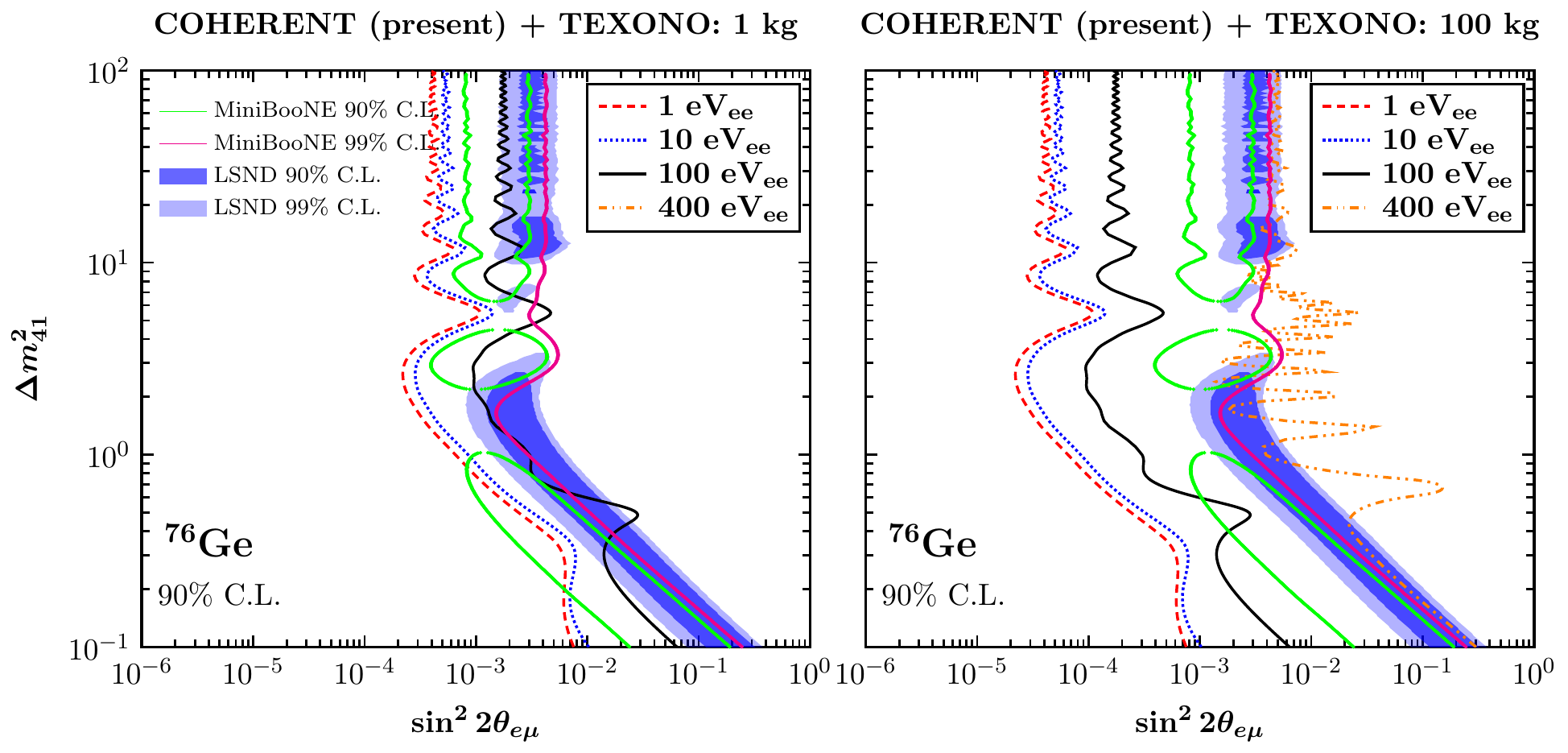}\\
\includegraphics[width= 0.8 \linewidth]{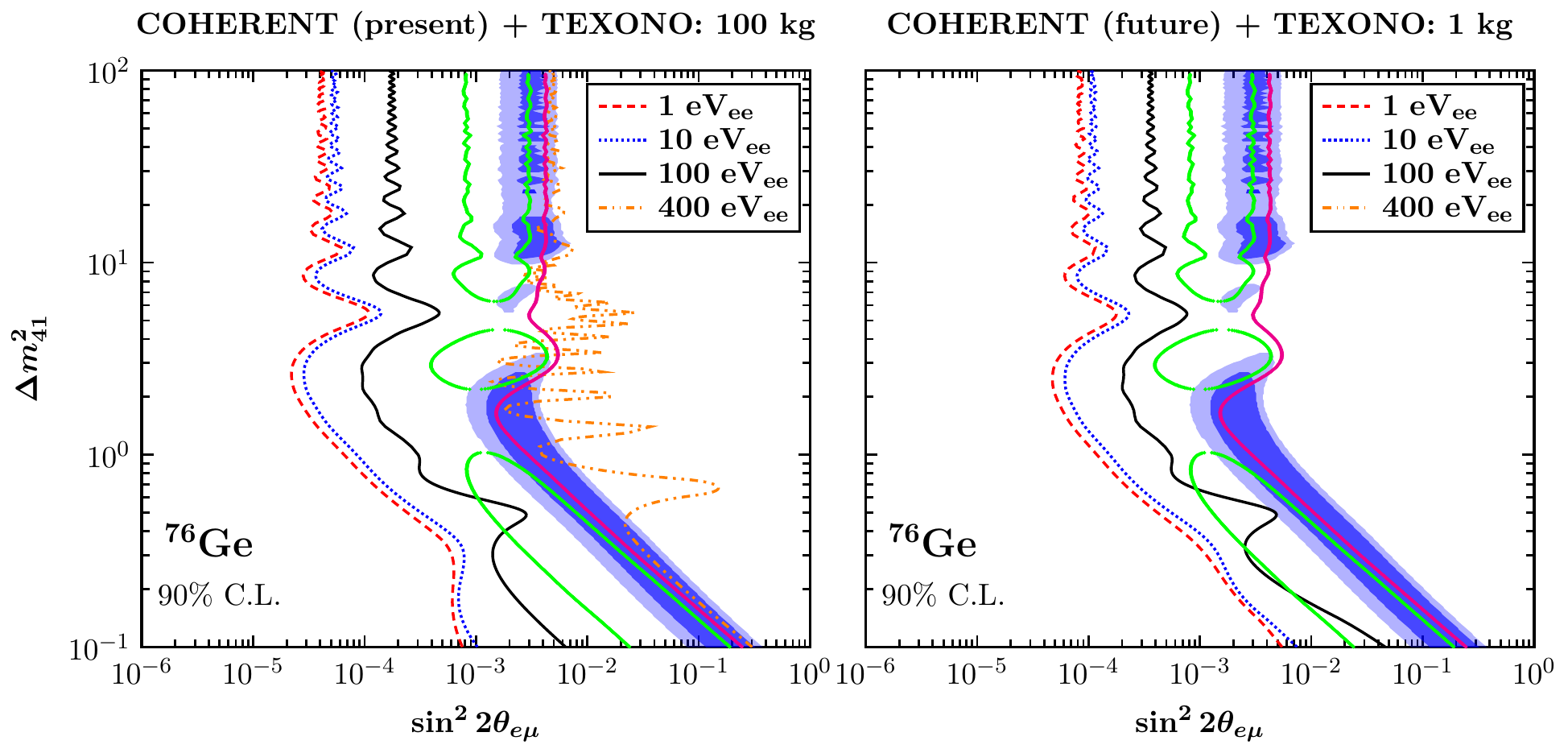}\\
\caption{Same as Fig.~\ref{fig.theta_emu.sys}, considering different setups for TEXONO and COHERENT and neglecting systematic uncertainties and backgrounds. For COHERENT, we are assuming $^{76}$Ge as target material.} 
\label{fig.theta_emu}
\end{figure*}
%%%%%%%%%%%%%%%%%%%%%%%%%%%%%%%%%%%%%%%%%%%%%%%%%%%%%%%%%%%%%%%%%%%%%%%%%%%%
%

Our present results indicate clearly that a dedicated experiment
searching for CE$\nu$NS has also satisfactory capabilities to probe
sterile neutrinos. 
For the case of the TEXONO experiment, the lack of $\bar{\nu}_e$
disappearance results in the sensitivity regions are depicted in the
top panel of Fig.~\ref{fig.sterile.sys} after one year of data taking
time by considering two extreme possibilities for the detector mass
(e.g. 1~kg and 100~kg).  In each case the assumed experimental setup
consists of a $^{76}$Ge detector with a $100~\mathrm{eV_{ee}}$
threshold and a background level of 1~cpd. For comparison purposes,
apart from the typical $\sigma_{\mathrm{sys}}=10\%$, two additional
more optimistic possibilities of the systematic error are also taken
into account: $\sigma_{\mathrm{sys}}=5\%$ and
$\sigma_{\mathrm{sys}}=1\%$. Moreover, we also explore a more
  conservative scenario by assuming a 10~cpd background level which
  shows differences only for the case of 1\% systematic error.  On
the other hand, in the bottom panel of Fig.~\ref{fig.sterile.sys} by
neglecting systematic errors and background events, the results are
also illustrated for two different values of the $^{76}$Ge target mass
(1~kg and 100~kg) and four possible energy thresholds
($1~\mathrm{eV_{ee}}$, $10~\mathrm{eV_{ee}}$, $100~\mathrm{eV_{ee}}$,
$400~\mathrm{eV_{ee}}$). For these thresholds, by using
Eq.(\ref{eq:Qu}) the corresponding quenching factors become
$\mathcal{Q}u=(0.12, 0.16, 0.23, 0.27)$ leading to nuclear recoil
thresholds ($8~\mathrm{eV_{nr}}$, $60~\mathrm{eV_{nr}}$,
$442~\mathrm{eV_{nr}}$, $1472~\mathrm{eV_{nr}}$), respectively. Note,
that this is also consistent with the choice $\mathcal{Q}_f=0.20-0.25$
employed in our previous work where we only considered a
$100~\mathrm{eV_{ee}}$ threshold~\cite{Kosmas:2015vsa}. We furthermore
note that, by assuming a threshold as high as
$T_{\mathrm{thres}}=400~\mathrm{eV_{ee}}$, the results indicate that
TEXONO has no sensitivity to sterile parameters for the case of 1~kg
$^{76}$Ge detector mass. One also sees that large values of
$\sin^2 2 \theta_{ee}$ would be ruled out by the exclusion curves, in
agreement with the results of
Refs.~\cite{Mention:2011rk,Giunti:2011gz}.  In addition, as stated in
Ref.~\cite{Giunti:2015mwa}, the requirement of large
$\vert U_{e1} \vert^2 + \vert U_{e2} \vert^2$ for solar neutrino
oscillations, implies that values of $\vert U_{e4} \vert^2$ close to
unity are excluded. Therefore, for small $\sin^2 2 \theta_{ee}$ one
has
\begin{equation}
\sin^2 2 \theta_{ee} \simeq 4 \vert U_{e4} \vert^2 \, .
\label{sin_e-e}
\end{equation}
which satisfies the general expectation that the fourth generation
massive neutrino is mostly sterile.

At this point we turn our attention on the capability of the COHERENT experiment~\cite{Akimov:2015nza} at the SNS, Oak Ridge, to probe the sterile neutrino parameters (for a comprehensive analysis, see also Ref.~\cite{Anderson:2012pn}).  Although SNS experiments in general involve both $U_{e4}$ and $U_{\mu 4}$, here we concentrate just on the latter since COHERENT is optimized to record muonic neutrino beams~\cite{Kosmas:2015vsa}. Focusing on  various promising nuclear targets at the SNS, in Fig.~\ref{fig.sterile-COHERENT.sys} we illustrate the expected sensitivity of the COHERENT to sterile neutrino parameters in the $(\sin^2 2 \theta_{\mu \mu }, \, \Delta m_{41}^2)$ plane, by assuming only the $\bar{\nu}_{\mu}$ component of the delayed neutrino flux. For the sake of comparison, the bottom panel of Fig.~\ref{fig.sterile-COHERENT.sys} shows the corresponding sensitivity by neglecting systematic uncertainties and background events for the case of $^{76}$Ge. The obtained results, in conjunction with the large values of
$\vert U_{\mu 1} \vert^2 + \vert U_{\mu 2} \vert^2 + \vert U_{\mu 3}
\vert^2 $ that are indicated by atmospheric neutrino
data~\cite{Maltoni:2007zf}, imply small values of
$\vert U_{\mu 4} \vert^2$.  Then, similarly to reactor neutrino
experiments, one may write
\begin{equation}
\sin^2 2 \theta_{\mu \mu } \simeq 4 \vert U_{\mu 4} \vert^2 \, .
\label{sin_mu-mu}
\end{equation}

Furthermore, a combination of Eq.(\ref{sin_ab}) with
Eqs.(\ref{sin_e-e}) and (\ref{sin_mu-mu}) yields the
appearance-disappearance constraint~\cite{Bilenky:1996rw}
\begin{equation}
\sin^2 2\theta_{e \mu} = \frac{1}{4} \sin^2 2\theta_{ee} \sin^2 2\theta_{\mu \mu} \, ,
\label{eq:combined-constraint-sterile}
\end{equation}
which implies that $\sin^2 2 \theta_{e \mu}$ is doubly suppressed for
small values of $\sin^2 2 \theta_{e e}$ and
$\sin^2 2 \theta_{\mu \mu}$. From the corresponding exclusion curve in
Fig.~\ref{fig.theta_emu.sys} by assuming various nuclear targets and
the previously described systematic uncertainties and backgrounds as
well as in Fig.~\ref{fig.theta_emu} for zero background events and
statistical errors only, we find that a combined analysis leads to a
high sensitivity for sterile neutrino searches. Confronting the present results with the respective allowed regions by LSND~\cite{Aguilar:2001ty}  and MiniBooNE~\cite{AguilarArevalo:2010wv, Aguilar-Arevalo:2012fmn,Aguilar-Arevalo:2013pmq} (see Figs. 7 and 8), we conclude that the expected sensitivity from CE$\nu$NS has the potential to probe them, especially after the future upgrade
of COHERENT and TEXONO.
These results are also competitive with recent sterile neutrino fits
obtained from global analyses of SBL neutrino oscillation
searches~\cite{Kopp:2011qd,Giunti:2011gz}.

%%%%%%%%%%%%%%%%%%%%%%%%%%%%%%%%%%%%%%%%%%%%%%%%%%%%%%%%%%%%%%%%%%%%%%%%%%%%%%
\section{Conclusions}
\label{sect:conclusions}
We have examined the potential of short-baseline coherent elastic
  neutrino-nucleus scattering experiments to probe effects
  associated to light sterile neutrinos. For definiteness we have
  focused on the normal (3+1) neutrino mass scheme.
We have found that the planned TEXONO and COHERENT experiments offer
good prospects of providing key information concerning the existence
of light sterile neutrinos. From our present results we conclude that
dedicated low-energy neutrino experiments looking for CE$\nu$NS events
could be complementary to charged-current appearance and disappearance
searches.
We have also verified that, by employing high-purity Germanium
detectors with sub-keV thresholds, better sensitivities can be reached
on the sterile neutrino mixing parameters. Such measurements would
provide a deeper understanding of neutrino interactions over a very
wide energy range and could possibly provide evidence for new
physics in the lepton sector.

%%%%%%%%%%%%%%%%%%%%%%%%%%%%%%%%%%%%%%%%%%%%%%%%%%%%%%%%%%%%%%%%%%%%%%%%%%%%%
\section*{Acknowledgements}

Work supported by MINECO grants FPA2014-58183-P, Multidark
CSD2009-00064, SEV-2014-0398, and the PROMETEOII/2014/084 grant from
Generalitat Valenciana. M. T. is also supported by the grant GV2016-142 (Generalitat Valenciana) and by a Ram\'{o}n y Cajal contract (MINECO). 
One of us, DKP, wishes to thank Prof. O. Miranda for stimulating
discussions and Dr. R. Fonseca for technical assistance. JWFV
acknowledges Prof. W.C. Louis for providing relevant experimental
data.

%%%%%%%%%%%%%%%%%%%%%%%%%%%%%%%%%%%%%%%%%%%%%%%%%%%%%%%%%%%%%%%%%%%%%%%%%%%%%
\section*{Note Added}

Recently, the COHERENT collaboration announced the observation of CE$\nu$NS for the first time at a 6.7~sigma confidence level, using a low-background, 14.6~kg
CsI[Na] scintillator~\cite{Akimov:2017ade}. This enhances the
significance of our results related to the COHERENT experiment.

%%%%%%%%%%%%%%%%%%%%%%%%%%%%%%%%%%%%%%%%%%%%%%%%%%%%%%%%%%%%%%%%%%%%%%%%%%
%\appendix


%\section*{References}
%\bibliographystyle{JHEP}   %if you use JHEP.bst
%\bibliographystyle{apsrev4-1}
%\bibliography{dimitris-2,newrefs,merged,references.bib,merged_Valle}

\end{document}